\documentclass[]{aa}

\usepackage{graphicx,natbib}
\usepackage{subfigure}
\usepackage{amsmath,amssymb}

\usepackage{txfonts}
\usepackage{pifont}

\usepackage{booktabs}

\usepackage{hyperref}

\hypersetup{
   colorlinks=true,
   citecolor=blue,
   linkcolor=blue,
   urlcolor=black
   }

\def\Rgp{R^{\star}}

\def\co2{CO$_2$}
\def\h2o{H$_2$O}
\def\ch4{CH$_4$}
\def\N2{N$_2$}
\def\nh3{NH$_3$}

\def\grav{g}
\def\Fs{F_{\star}}

\def\press{p}
\def\hp{\pi}
\def\prcb{\press_\mathrm{RCB}}
\def\temp{T}
\def\teta{\theta}
\def\tetav{\theta_\mathrm{v}}
\def\Teta{\Theta}

\def\time{t}
\def\q{q}
\def\qi{\q_{\i}}
\def\qint{\q_\mathrm{int}}
\def\kzz{K_{zz}}
\def\kzzmin{K_{zz}^{\mathrm{min}}}
\def\kzzmax{K_{zz}^{\mathrm{max}}}
\def\tautra{\tau_\mathrm{tra}}
\def\rtra{r_\mathrm{tra}}

\def\x{x}
\def\y{y}
\def\z{z}
\def\geop{\phi}

\def\dry{d}
\def\vap{v}
\def\cond{c}

\def\pzero{\press_0}
\def\hpd{\hp_{\dry}}

\def\hpm{\hp}
\def\hpt{\hp^{t}}
\def\hpds{\hpd^{s}}
\def\hpms{\hpm^{s}}

\def\r{r}

\def\rv{\r_{\vap}}

\def\ri{\r_{\i}}
\def\rR{\chi}
\def\rRi{\rR_{\i}}
\def\q{q}
\def\qd{\q_{\dry}}
\def\qv{\q_{\vap}}

\def\qcri{q_\mathrm{cri}}

\def\cp{c_\press}

\def\R{R}
\def\Rd{\R_{\dry}}
\def\Rv{\R_{\vap}}

\def\M{M}
\def\Md{\M_{\dry}}
\def\Mv{\M_{\vap}}
\def\Latent{L}

\def\massd{\mu_{\dry}}
\def\etad{\eta_{\dry}}
\def\etadp{\dot{\eta}_{\dry}}
\def\Etadp{\Omega}
\def\vol{\alpha}
\def\vold{\vol_{\dry}}

\def\volm{\vol}

\def\w{w}
\def\W{W}
\def\vv{\mathbf{v}}
\def\vvh{\bar{\mathbf{v}}}
\def\vV{\mathbf{V}}
\def\vVh{\mathbf{\bar{V}}}

\def\xforce{F_{\x}}
\def\yforce{F_{\y}}
\def\zforce{F_{\z}}
\def\Q{Q}
\def\Qi{Q_\i}
\def\Flux{F}
\def\Fluxi{\Flux_\i}
\def\Source{S}
\def\Sourcei{\Source_{\i}}
\def\Finc{F_{inc}}
\def\alb{A}

\def\d{\mathrm{d}}
\def\i{\mathrm{i}}

\def\l{l}

\newcommand{\balign}[1]{
\begin{align}
#1
\end{align}}

\newcommand{\eq}[1]{Eq.\,(\ref{#1})}

\newcommand{\fig}[1]{Fig.\,\ref{#1}}
\newcommand{\figs}[2]{Figs.\,\ref{#1} and \ref{#2}}
\newcommand{\sect}[1]{Sect.\,\ref{#1}}

\newcommand{\app}[1]{Appendix\,\ref{#1}}
\newcommand{\tab}[1]{Table\,\ref{#1}}

\newcommand{\dpartial}[2]{\frac{\partial #1}{\partial #2}}
\newcommand{\dpar}[2]{\partial_{#2}#1}

\titlerunning{Revisiting the atmospheric structure of temperate Neptunes}
\authorrunning{Leconte et al.}

\begin{document}



\title{A 3D picture of moist-convection inhibition\\ in hydrogen-rich atmospheres: Implications for K2-18\,b}

\author{J\'er\'emy Leconte\inst{1}, Aymeric Spiga\inst{2}, No\'e Cl\'ement\inst{1}, Sandrine Guerlet\inst{2,3}, Franck Selsis\inst{1}, Gwena\"el Milcareck\inst{2}, Thibault Cavali\'e\inst{1,3}, Rapha\"el
Moreno\inst{3}, Emmanuel Lellouch\inst{3}, \'Oscar Carri\'on-Gonz\'alez\inst{3}, Benjamin Charnay\inst{3}, Maxence Lef\`evre\inst{4}}

\institute{
Laboratoire d'astrophysique de Bordeaux, Univ. Bordeaux, CNRS, B18N, all\'ee Geoffroy Saint-Hilaire, 33615 Pessac, France
\and
Laboratoire de M\'et\'eorologie Dynamique (IPSL), Sorbonne Universit\'e, Centre National de la Recherche Scientifique, \'Ecole Polytechnique, \'Ecole Normale Sup\'erieure, Paris, France
\and
LESIA, Observatoire de Paris, Universit\'e PSL, CNRS, Sorbonne Universit\'e, Universit\'e de Paris, 5 place Jules Janssen, 92195 Meudon, France
\and
LATMOS/IPSL, Sorbonne Université, UVSQ Universit\'e Paris-Saclay,  CNRS, Paris, France
}

\date{Submitted 14 December 2023 / Accepted 9 January 2024}

\offprints{jeremy.leconte@u-bordeaux.fr}

\abstract{
While small, Neptune-like planets are among the most abundant exoplanets, our understanding of their atmospheric structure and dynamics remains sparse. In particular, many unknowns remain on the way moist convection works in these atmospheres where condensable species are heavier than the non-condensable background gas. While it has been predicted that moist convection could shut-down above some threshold abundance of these condensable species, this prediction is based on simple linear analysis and relies on some strong assumptions on the saturation of the atmosphere. To investigate this issue, we develop a 3D cloud resolving model for hydrogen-dominated atmospheres with large amounts of condensable species and apply this model to a prototypical temperate Neptune-like planet --- K2-18\,b. Our model confirms the shut-down of moist convection above a critical abundance of condensable vapor and the onset of a stably stratified layer in the atmosphere of such planets, leading to much hotter deep atmospheres and interiors. Our 3D simulations further provide quantitative estimates of the turbulent mixing in this stable layer, which is a key driver of the cycling of condensables in the atmosphere. This allows us to build a very simple, yet realistic 1D model that captures the most salient features of the structure of Neptune-like atmospheres. Our qualitative findings on the behavior of moist convection in hydrogen atmospheres go beyond temperate planets and should also apply to the regions where iron and silicates condense in the deep interior of hydrogen-dominated planets. Finally, we use our model to investigate the likelihood of a liquid ocean beneath a H$_2$ dominated atmosphere on K2-18\,b. We find that the planet would need to have a very high albedo ($\alb>0.5-0.6$) to sustain a liquid ocean. However, due to the spectral type of the star, the amount of aerosol scattering that would be needed to provide such a high albedo is inconsistent with the latest observational data. 
}


\maketitle

\section{Introduction}

Convection, the process by which an unstably stratified fluid transports energy upward to restore neutrality, shapes the thermal structure of all the deep planetary atmospheres in our Solar System. Inherently caused by the difference in the radiative opacity of the atmosphere between the wavelengths at which it is heated by the star and those at which it can cool by thermal radiation, convection usually happens wherever radiative processes are insufficient to carry energy out. This leads to atmospheres consisting of two main layers: a deep troposphere overlain by a stratosphere. This structure stems from such basic principles that it is envisioned to hold true on most exoplanets with a substantial atmosphere \citep{RC14}, although the intense irradiation they receive can mean that the stable radiative zone can extend very deep \citep{GS02}. 

But for this balance to hold, convection itself needs to be able to develop, for it can be hindered by other dynamical processes. For example, when there is a compositional gradient in the atmosphere - whether it is caused by condensation or chemical reactions - the resulting mean molecular weight gradient in the gas can affect the thermal gradient needed to initiate convection \citep{NTI00, Gar18, DPT21, HP23}. In a more extreme fashion, when the atmosphere contains enough of a condensable species (hereafter referred to as \textit{vapor}) that is heavier than the non-condensable background gas (hereafter referred to as \textit{air}), condensation can completely suppress convection, whatever the thermal gradient \citep{Gui95, LSH17, MGS22}. Intuitively, this is due to the fact that when we consider saturated moist air following the Clausius-Clapeyron law, there is a threshold specific concentration of condensable vapor, 
\balign{\qcri\equiv \frac{1}{\Mv-\Md}\frac{\Rgp \temp}{\Latent}, \label{qcri}
} above which a change in vapor abundance due to condensation affects buoyancy more (and in the opposite direction) than the temperature change that led to it. In this equation, $\Mv$ and $\Md$ are the molar masses for the vapor and dry air respectively, $\Latent$ is the specific latent heat, $\temp$ the temperature, and $\Rgp$ the universal molar gas constant. Above that threshold, the effective thermal expansion coefficient of the fluid becomes negative \citep{MGS22}. A super-adiabatic region thus becomes perfectly stable to convection because parcels of fluid become denser than their environment as they rise. 

\cite{LSH17} hypothesized that in hydrogen-dominated atmospheres, a stable layer would form near the cloud deck of any species for which the deep abundance would be higher than the critical concentration~$\qcri$. This critical concentration is probably reached for water in Saturn \citep{LI15}, and for water and methane in Uranus, Neptune, and most exo-Neptunes. \citet{MGS22} and \citet{MS22} independently extended this argument to iron at the core-envelope boundary in small neptune-like exoplanets. The main consequence of this stable, super-adiabatic layer is that the deep atmosphere and interior can be much hotter than predicted by standard models for the same effective temperature (hence the same temperature of the upper troposphere and stratosphere), with implications for the past evolution and chemistry of our ice giants \citep{CVS17,CVM20,MS21}.

However, the process of convection inhibition has been mostly studied analytically with unidimensional arguments, and many caveats remain:
\begin{itemize}
\item[$\bullet$] The analytical theory of \cite{LSH17} assumed a fully-saturated medium, whereas we know that on Earth, moist convective regions are all but fully-saturated. Would a stable layer really form in a realistic, subsaturated atmosphere? 
\item[$\bullet$] If the stable layer is indeed devoid of any turbulence, as has been predicted, it should prevent any upward transport of vapor, entailing a dry upper troposphere without moist convection. However, this contrasts with the observation of moist convective systems in the atmosphere of ice giants. Why is that? 
\item[$\bullet$] How does convection inhibition affect the atmosphere of neptune-like exoplanets and is this observable with our current methods and observatories?
\end{itemize}

To answer these questions, we develop a 3D cloud-resolving model that is able to simulate hydrogen-dominated atmospheres with large amounts of any condensable volatile (see \sect{sec:3dmodel}). Because the atmosphere of Solar System ice giants have radiative timescales on the order of decades to centuries, we focus on simulating the atmosphere of a prototypical temperate sub-neptune: K2-18b \citep{CAD17}. While enabling the equilibration of the simulation with reasonable computing resources, this also helps us shed light on the atmospheric structure of such temperate, hydrogen-dominated planets (\sect{sec:simulation}). We argue that many features exhibited by convection in this regime should be fairly general and apply to other systems with moist-convection inhibition. In particular, we will show that, while being more realistic, the 3D simulations largely support the thermal structure envisioned in the earlier study of \cite{LSH17}. In addition, a detailed study of the rich energy and volatile cycles in the 3D simulations allows us to build a much more consistent picture of these atmospheres and to develop a brand new 1D model that can be used to explore their diversity (\sect{sec:1dmodel}). Finally, in \sect{sec:observability}, we discuss how the chemical composition of temperate Neptune-like planets is affected by convection inhibition, potentially acting as a tracer of this process.

Interestingly, during the writing of this manuscript, new near- and mid-infrared JWST transit observations of K2-18\,b have been released \citep{MSC23}. One of the most intriguing feature of these observations is the potential non-detection of NH$_3$ that has been interpreted as the sign of a shallow atmosphere overlying a liquid ocean. Indeed, if a deep atmosphere were present, the thermochemical conditions at the bottom of such an atmosphere should replenish the upper atmosphere in ammonia \citep{HDS21}. Although this non-detection will probably need to be confirmed when our knowledge of instrumental systematics improve, we have tried to quantify in more details whether or not the presence of a liquid ocean can be consistent with both these new observations and our improved knowledge of the thermal structure of these temperate sub-neptunes. For that purpose, in \sect{sec:runaway}, we use our updated 1D model to put limits on the conditions necessary to sustain a liquid ocean on K2-18\,b and show that this requires planetary Bond albedos in excess of 0.5-0.6, that are relatively hard to achieve around a late-type star. We further demonstrate that the kind of aerosol properties that we need to create such a high albedo are incompatible with the current transit observations that exhibit relatively deep near-infrared methane bands.

\section{3D hydrodynamical cloud-resolving model}\label{sec:3dmodel}

The mechanism for moist-convection inhibition rests on a fairly basic principle: when the mean molecular weight of the vapor is larger than the one of the air, a parcel of fluid becomes denser as its vapor concentration increases. The ingredients for convection inhibition are therefore inherently present in the basic equations of the hydrodynamics of a moist atmosphere. However, in many terrestrial atmospheres, the condensable species are found in trace amounts and neglecting their contribution to the density of the air and to the global mass of the atmosphere is a rather valid assumption, used in many numerical models. Such models thus cannot exhibit convection inhibition.

Fortunately, the "trace gas" assumption is challenged when modeling strong convective events in hot and moist regions of the Earth---even if on Earth, the presence of water vapor facilitates convection as it is lighter than molecular nitrogen. For this reason, many small-scale, non-hydrostatic models of the atmosphere -- the so-called cloud-resolving models -- incorporate the mass-loading effect of water vapor \citep{BF02,SKD19}. 

To model the convection in hydrogen-rich atmospheres, we thus used the dynamical core of the 4th version of the Weather Research and Forecast model (hereafter WRF V4) described in \citet{SKD19} that we coupled with the physical parametrizations of the LMD Generic Planetary Climate Model (Generic PCM, the former Generic LMD GCM; \citealt{WFS11,LFC13}). The following sections describe the details of each code and their coupling as well as the specific developments and the numerical setup used. 

\subsection{Equations}

We 
simulate the compressible, non-hydrostatic Euler equations using the philosophy described in \citet{Lap92} that uses a terrain-following hydrostatic-pressure vertical coordinate defined as:
\balign{\etad \equiv \frac{\hpd-\hpt}{\hpds-\hpt}\equiv \frac{\hpd-\hpt}{\massd},}
where $\hpt$ and $\hpds$ are the dry hydrostatic pressure at the top of the model and at the surface (it is assumed that there is negligible moisture above the model top), and dry mass in the model column is $\massd \equiv \hpds-\hpt$. 

The hydrostatic pressure, $\hp$, is the local pressure the fluid would have if in hydrostatic equilibrium with the same mass above the point considered. For a constant gravity, it is equal to the mass of fluid above a given point divided by gravity.
Defining the geopotential, $\geop\equiv \grav \z$, and the specific volume of the fluid, $\vol$, one can write the equivalent of the hydrostatic equation
\balign{
\dpartial{\geop}{\hp}=- \vol,
}
except this equation is valid even when there is no equilibrium. 
When there is moisture in the atmosphere, we define $\hpd$ as the mass of dry air above a given point divided by $\grav$. It can be easily shown that
\balign{\label{defhp}
\dpartial{\hpm}{\hpd}=\frac{\vold}{\volm},
}
so that
\balign{\label{defhp}
\dpartial{\geop}{\hpd}=- \vold.
}

The mass-weighted, prognostic atmospheric variables in the flux-form dynamical equations are 
\balign{\vV \equiv (U,V,\W) \equiv \massd \vv, \ \Etadp \equiv \massd \etadp, \  \Teta\equiv \massd \teta, \ \rRi \equiv \massd \ri,}
and the geopotential~$\geop$ that is not written in flux form as it is not a conserved quantity. The usual velocities are denoted $\vv=(u,v,w)$ and we define the horizontal velocity for sake of compacity $\vVh \equiv \massd \vvh \equiv \massd (u,v)$. $\teta \equiv \temp (\pzero/\press )^{\R/\cp}$ is the potential temperature, $\pzero$ is an arbitrary reference pressure, $\R$ the specific constant of the gas, and $\ri$ is the mass mixing ratio of the various tracer species with respect to dry air (in particular water vapor with index $\vap$ and condensed water with index $\cond$). For the sake of compactness of notation, the specific concentration of the tracers, $\qi = \ri / (1+\rv)$, will sometimes be used when appropriate.

Adapting the equations of \citet{Kas74} and \citet{Lap92} to a moist atmosphere yields the following prognostic equations in flux form:
\balign{
\dpar{U}{\time}+   (\nabla \cdot \vV  u)  -\massd \vol\  \dpar{\press}{x}+(\vol/\vold)\ \dpar{\press}{\etad}\ \dpar{\phi}{x} &=\massd \xforce,\\
\dpar{V}{\time}+   (\nabla \cdot \vV  v)  -\massd \vol\  \dpar{\press}{y}+(\vol/\vold)\ \dpar{\press}{\etad}\ \dpar{\phi}{x} &=\massd \yforce,\\
\dpar{\W}{\time}+   (\nabla \cdot \vV  \w)  -\grav\left[(\vol/\vold)\ \dpar{\press}{\etad} -\massd \right] &=\massd \zforce, \\
 \dpar{\massd}{\time} + \left(\nabla\cdot \vV \right) &=0, \\
\dpar{\geop}{\time} + \frac{1}{\massd} \left[ \left(\vV \cdot \nabla \geop \right)- \grav \W \right]&=0, \\
\dpar{\Teta}{\time} + \left(\vV \cdot \nabla \teta \right)&=\massd \frac{\teta}{\temp}\frac{\Q}{\cp},  \label{eq_entropy}\\
\dpar{\rRi}{\time} + \left(\vV \cdot \nabla \ri \right)&=\massd \Sourcei,
}
where the various differential operators are defined as follows
\balign{
& \left(\nabla\cdot \vV \right) \equiv\dpar{U}{x} +\dpar{V}{y} + \dpar{\Etadp}{\etad} ,\\
& \left(\nabla\cdot \vV a \right) \equiv \dpar{Ua}{x} +\dpar{Va}{y} + \dpar{\Etadp a}{\etad}, \\
& \left(\vV \cdot \nabla a \right) \equiv U\dpar{a}{x} +V\dpar{a}{y} +\Etadp \dpar{ a}{\etad}.
 }
The remaining quantities are $(\xforce, \yforce, \zforce)$ the external forces per unit mass exerted on the fluid, $\Q$ the specific, diabatic heating rate, and $\Sourcei$ the specific source/sink rates for tracer $\i$ (in our case, the vapor and condensed phases).

Our set of equations differ from either \citet{DPT21} or \citet{HP23} who chose to solve an equation for the conservation of energy directly, whereas we chose to formulate the problem with a potential temperature. The effect of this numerical choice should probably be investigated in more depth by simulating the same setup with these different codes.

To close the system, we provide an equation of state for the ideal mixture of vapor and dry gas that writes
\balign{
\press = \pzero \left(\Rd \teta \left[1+(\Rv/\Rd) \rv \right] / \pzero \vold \right)^{\gamma},
}
where $\Rv$ and $\Rd$ are the specific gas constant of the vapor and dry air, and $\gamma=1/(1-\R/\cp)$ is kept constant. Notice that the ratio $\R/\cp$ for the whole gas needs to be constant to express the entropy equation in the form of an equation on a potential temperature (\eq{eq_entropy}). However, because $\R=\qd \Rd +\qv \Rv$ will change with the vapor concentration, the $\cp$ used in the right-hand side of \eq{eq_entropy} must be varied accordingly, i.e. following $\cp=\qd c_{p,d} +\qv c_{p,v}$. This corresponds to a situation where the vapor and the dry gas would have the same molar heat capacity. 

Apart from its effect on the equation of state, the vapor and condensates affect the dynamics through the $\vold/\vol \equiv 1 + \sum_{\i} \ri$ terms that account for the mass loading effect of both phases.

\subsection{Dynamical core and numerical implementation}

To solve these equations, we use the 4th version of the Weather Research and Forecast model (hereafter WRF V4) described in \citet{SKD19}. We refer the reader to this technical note for all the details of the general numerical implementation and of the various schemes and options mentioned hereafter. 

Our baseline simulations are performed on a 64x64x256 grid with a 2\,km horizontal resolution. The physical vertical resolution is variable as the grid is based on fixed $\etad$ levels but varies around 400\,m in most of the domain with thinner layers near the surface. Despite the very large scale height of the modeled atmospheres compared to Earth, test simulations with coarser vertical grids seem to indicate that such thin layers are necessary to satisfactorily resolve the convection. The horizontal boundaries are periodic, the surface is a rigid lid and the top of the model is a fixed pressure boundary around 3000\,Pa. A damping layer extends over the top 20\,km to damp upward propagating gravity waves following the vertical-velocity implicit Rayleigh damping scheme.

The dynamical timestep was fixed for each simulation to fulfill the CFL stability condition and was on the order of 5\,s. Physical parametrization described in the next section are called every 75\,s and radiative transfer calculations are performed once every ten minutes. 

As one of the goals of the study is to identify the dynamical processes that can transport energy and tracers in the modeled atmospheres, we turn off all the parametrizations of subgrid-scale turbulent transport usually included in cloud-resolving models. This enables us to know that any transport observed in the simulations is the result of resolved dynamical motions and not due to ad-hoc parametrizations \citep{PSL13}. 

The various tracers are transported by the dynamics with the 5th order \textit{monotonic} advection scheme recently implemented in WRF V4. This proved crucial in providing physical results in the presence of a sharp vapor gradient arising in the stable layer above convective motions. In particular, using only the positive definite scheme (as implemented in WRF V3) creates a spurious local minimum of vapor concentration above the stable layer that dried the troposphere above, suppressing moist convection there. This is in accordance with the known tendency of non-monotonic advection schemes to create spurious minima around sharp edges \citep{Ska06}, but here the manifestation goes well beyond a small inaccuracy, warranting the use of a monotonic scheme.

\subsection{Physical parameterisations} \label{sec:phys_param}

The various physical source terms in the equations of motion, in particular the diabatic heating and the tracer source terms, $\Q$ and $\Sourcei$ respectively, are computed in each column of the model using the physical parametrizations from the Generic PCM. This strategy of coupling the WRF dynamical core with physical parameterizations for planetary atmospheres 
has been developed
for martian applications in \citet{Spi09} and was later extended to Venus \citep{LLS18} and temperate exoplanets \citep{LTP21}.

In the present study, the coupling has been improved in several ways:
\begin{itemize}
\item We transitioned toward the 4th version of WRF to take advantage of the new features of the dynamical core. The inclusion of a monotonic advection scheme, in particular, was instrumental in properly modeling the steep gradients in the stable layer. 
\item As we model water vapor rich atmospheres, a particular attention has been devoted to improving the water conservation in the interface between the dynamical core that uses tracer mixing ratios ($\ri$) and the physical parametrizations in PCM that use tracer specific concentration with respect to the whole gas ($\qi = \ri / (1+\rv)$) -- the difference between those two variables, which was neglected in previous versions, is no longer negligible in the case studied here. 
\item As the usual PCM global climate model assumes hydrostatic equilibrium, the total thermodynamic pressure at level interfaces was used to calculate the mass of gas in model layers. This led to a non-conservation of the total mass of the atmosphere. The hydrostatic pressure ($\hp$) provided by the WRF dynamical core is now used instead. 
\end{itemize}

In terms of physical processes, many parametrizations in the physical part of the Generic PCM have been developed to be used with a hydrostatic, global dynamical core: dry and moist convective adjustment, subgrid-scale humidity distribution, etc. These are turned-off when running the Generic PCM in a cloud-resolving mode because these processes are resolved in the simulations. The main remaining parametrizations are the radiative transfer and the formation and evaporation of precipitations. 

For the radiative transfer, use the generic two-stream module of the PCM with correlated-$k$ coefficient tables. For these specific simulations, we used \texttt{Exo\_k} \citep{Lec21} to bin-down and combine the correlated-$k$ table opacities from \citet{BCB21} into tables with 21 channels for the thermal emission and 9 for the stellar radiation with 25 pressure points uniformly distributed in log $P$ between 0.1 and 10$^7$\,Pa, 15 temperature points, and 9 values of the water vapor mixing ratio that is variable in the simulations. The abundances of the other species depend on the chosen metallicity and will be detailed hereafter. The CIA for H$_2$-H$_2$, H$_2$-He are taken from the HITRAN database \citep{RGR11}. The  H$_2$O-H$_2$O and H$_2$O-air continua are based on MT-CKD 3.3 \citep{MPM12}. The stellar spectrum is approximated by a blackbody at the effective temperature of the star and the domain receives a constant insolation with an effective solar zenith angle of 60$^\circ$.

For the precipitations, we have opted for a very simple scheme to isolate the contribution of the dynamics and be able to later identify how more realistic microphysics impact the atmospheric structure. Parcels of air that are supersaturated are instantaneously brought back to liquid-vapor equilibrium iteratively to account for the latent heat effect. Condensates are assumed to precipitate whenever their mixing ratio exceeds a threshold that we keep arbitrarily small in this first study to avoid the radiative feedback of clouds. Precipitations fall instantaneously. Because the planets we model are envisioned to have a thick atmosphere that reaches deeper that what we can model, depositing the precipitations at the model surface would be rather unphysical. Consistently with our choice of a simple microphysics, we want to keep a physically motivated model without any free parameter. We thus assume that evaporation is inefficient until droplets reach the \textit{boiling} level where bubbles of vapor would form inside the droplets. It is indeed improbable that rain drops would fall much deeper than that, so that this is clearly an upper limit on the pressure of the reevaporation level. An other advantage of this scheme is that it should be conservative in the sense that it will favor the formation of unsaturated layers. This is very appropriate considering that we want to test whether saturation in the stable layer is required to suppress convection.

Because the WRF dynamical core takes into account the weight of the vapor, part of the thermal energy absorbed by the atmosphere is converted to potential energy when this vapor is transported aloft. This is reconverted back to heat when precipitations fall and dissipate their potential energy through friction \citep{FML06, DP16}. Therefore, to close the energy budget, we assume that falling precipitations reach their equilibrium velocity instantaneously so that the potential energy liberated by condensates crossing a layer is deposited directly in that layer (See \citet{FML06} for details).

To simulate the fact that the deep atmosphere can act as an infinite reservoir of vapor, the vapor concentration in the first simulation level is always restored to the imposed internal vapor concentration, $\qint$, at each physical timestep. We keep the bottom of the model below the reevaporation level and verify that this pseudo exchange of vapor with the interior is negligible when the simulation is equilibrated.

\subsection{Energy conservation}\label{sec:nrjcons}

We observe that there is a net deficit of thermal emission of the atmosphere compared to incoming radiation, even when the model is equilibrated. In our baseline simulation, this deficit is about 1-2\% of the incoming flux, which seems reasonable considering the level of accuracy sought. 
Yet, we here try to identify what are the possible sources of such energy losses and to identify possible areas of improvement for the model in the future.

Let us first remind the reader that, although our equations do conserve energy in the dry gas regime \citep{Lap92}, our dynamical core is not formulated specifically in an energy-conserving way. This can be seen in \eq{eq_entropy}, which uses potential temperature (entropy) as its conserved variable. So some numerical losses of energy are inevitable. Apart from that, we think the various losses come from the following sources
\begin{itemize}
\item[$\bullet$] Gravity waves, which are launched by updrafts in the dry and moist convective zones carry away mechanical energy upward. When they reach the top sponge layer, this energy is dissipated without being reconverted to heat. 
\item[$\bullet$] To remove the well-known problem of the build-up of energy at the grid scale 
caused by
the turbulent cascade, the dynamical core implements various filters that are supposed to dissipate this energy. Here again, the dynamical core has not been designed to convert this dissipation back to heat, creating an energy sink.
\item[$\bullet$] Although using a potential temperature requires only the \textit{ratio} $\R/\cp$ to be constant, the demonstration of energy conservation in \citet{Lap92} relies on the specific heat capacity of the whole gas also being constant. This is impossible when the molar mass (hence the specific gas constant) of the dry gas and vapor are different and that vapor concentration is allowed to change. Even if we have taken into account the variation of $\cp$ in \eq{eq_entropy}, this variation should also entail a slight non-conservation, especially when the mixing ratio of vapor is not negligible.
\end{itemize}

\begin{table}[htbp]
\centering
\caption{Parameters used in the baseline 3D simulation.}
\begin{tabular}{llll}
\hline \hline
Surface gravity & $\grav$ &[m s$^{-2}$] & 12.41 \\
Average insolation & $\Fs$ &[W m$^{-2}$] & 175. \\
Deep vapor concentration & $\qint$ &[kg.kg$^{-1}$] & 0.45 \\
Specific heat capacity & $\cp$ &[J\,K$^{-1}$kg$^{-1}$] & 5470.\\
Molar mass of air & $\Md$ &[kg\,mol$^{-1}$] & 5.42$\times10^{-3}$ \\
Surface albedo & $A_\mathrm{s}$ && 0.\\
\hline \hline\\
\end{tabular}\label{tab:params}
\end{table}

\section{Atmospheric dynamics and thermal structure of temperate Neptunes}\label{sec:simulation}

\begin{figure*}[htb]  
\includegraphics[scale=0.82,trim = 0.2cm 0cm 0.cm 0cm, clip]{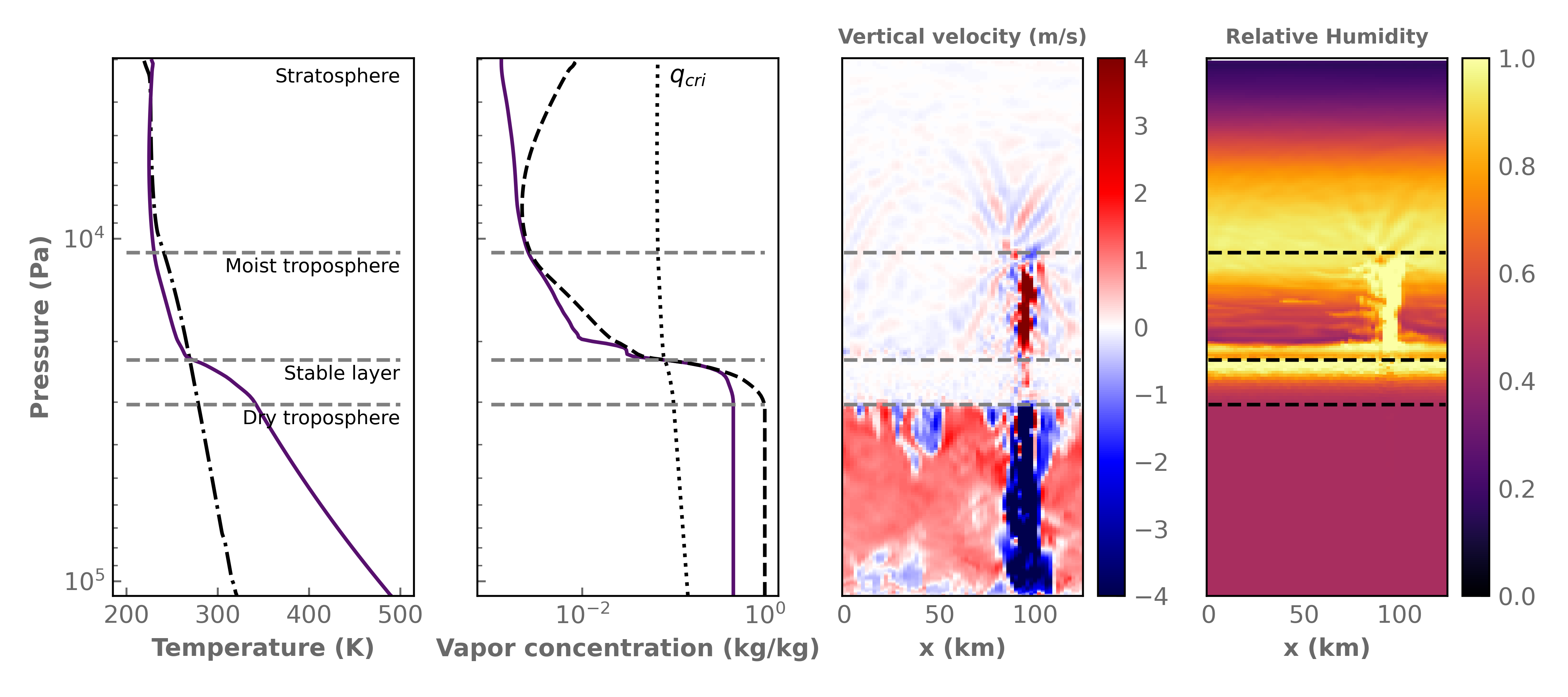}
\caption{Atmospheric structure of the baseline simulation. From left to right: Temperature, Water vapor specific concentration, vertical velocity (in m/s), and relative humidity. The two first panels show horizontal and temporal averages. The black, dash-dotted line in the first panel show the standard moist-adiabat profile for the same conditions. In the second panel, the black dotted and dashed lines show the value of the critical inhibition vapor concentration (\eq{qcri}) and the saturation concentration respectively. The two last panels show snapshots along vertical slices that go through a moist convective plume. From bottom to top, the atmosphere exhibits a dry troposphere, a stable layer where vertical motions are strongly suppressed, a moist troposphere, and a stratosphere. Horizontal dashed lines are plotted at the boundaries between these zones to allow an easier comparison of the altitudes of the various features. The rising moist plume (with maximal velocities of around 8\,m/s) is mirrored by a descending cold plume in the dry region (-15\,m/s) caused by the reevaporation of rains at the bottom of the stable layer.   }
\label{fig:structure_moist}
\end{figure*} 

We now present our simulations of a prototypical temperate Neptune-like exoplanet: K2-18b \citep{CAD17}. Indeed, compared to our Solar System ice giants, the radiative timescale of the atmosphere is on the order of days to months, instead of decades to centuries. Running a full cloud-resolving model to thermal equilibrium thus becomes feasible. Despite this choice, the very fine vertical discretization and the small dynamical step that it entails make these simulations relatively long and expensive to run: several months on 20 CPUs for several years of simulated time.

To limit the carbon budget of this study, we thus decided to severely limit the number of configurations that we simulated in 3D and use these to develop a 1D model to explore the parameter space as will be presented in \sect{sec:1dmodel}. In this section, we thus delineate the salient features of the thermal, compositional, and dynamical structure of these atmospheres that need to be incorporated in the simpler 1D model for it to be realistic.

\subsection{Simulation setup}

The parameters used for our baseline simulation are summarized in \tab{tab:params}. The atmospheric composition has been computed for a metallicity of 300$\times$ solar, which is consistent with the observations \citep{BCB21}. In principle, the precise atmospheric composition should depend on the temperature at depth, which itself depends on the modeled convection, hence on the composition through the mean molecular weight. We thus decided to use a simple approach to convert metallicity into molecular abundances: we assume that the quenching in our temperate planet will occur near 10\,bar and 1000\,K and use the chemical code of \citet{VCB20} to compute the abundances. With this approach, the volumic concentration of the main absorbers are $x_{H_2O}$=2.$\times 10^{-1}$, $x_{CO_2}$=1.5$\times 10^{-2}$, $x_{CH_4}$=7.4 $\times 10^{-2}$, $x_{CO}$=2.7$\times 10^{-2}$, $x_{NH_3}$=2.9$\times 10^{-4}$, $x_{He}$=1.5$\times 10^{-1}$, and $x_{H_2}$=5.2$\times 10^{-1}$. For water, this corresponds to a deep specific concentration $\qint=$0.45 kg/kg, but the concentration in each cell is traced and computed by the model. The other components are assumed to remain in same proportions everywhere and to form the so-called \textit{dry air}, with a molar mass of 5.42$\times10^{-3}$\,kg\,mol$^{-1}$.

Because K2-18b receives an insolation that is very close to the runaway greenhouse threshold (see discussion in \sect{sec:runaway}), the extreme climate sensitivity around this transition makes equilibration of the model very long for the observed insolation. For this reason, and since the goal of our study is to understand the general behavior of such atmospheres, we use an average insolation of 175\,W/m$^2$ to run our 3D simulations. This would correspond to an effective bond albedo of $\approx 0.5$, which allows us to be in the right regime to test the conclusions of \citet{MSC23} in \sect{sec:runaway}.

\subsection{Thermal structure}
The equilibrated atmospheric structure is depicted in \fig{fig:structure_moist}. One of our main findings is that, despite 
a more complex humidity distribution in our 3D simulations, we confirm the structure predicted by \cite{LSH17} using 1D idealized simulations: a stable layer forms between the level where $\q=\qcri$ and the dry troposphere below. This stable layer is clearly noticeable due to i) its superadiabatic thermal gradient and ii) its very low convective velocities. Indeed, no convective plume, either dry or moist, does penetrate this layer, so that there is no local source of gravity waves or turbulence. It can also be noted that the static stability in this layer is very high so that gravity waves have a much smaller vertical wavelength compared to the stratosphere.

A new feature of the 3D simulation is that it predicts the existence of a very thin, dry boundary layer between the stable layer below and the moist troposphere above. This is reminiscent of the surface boundary layer on Earth where 
small-scale dry plumes 
carry humidity from the surface to the condensation level where moist convection can occur.

As a comparison, we also show in \fig{fig:structure_moist} the thermal structure for a standard moist-adiabatic atmosphere with the same insolation and parameters (black dash-dotted curve). The tropopause exhibits similar pressure levels and temperatures in both cases, which is to be expected as they have the same flux to output and similar compositions. However, the temperature at depth varies dramatically---the moist adiabat being more than 150\,K colder at the 1\,bar level. The reason for this is twofold: first, the stable layer created a huge temperature jump in a narrow vertical region and, second, because the atmosphere below the stable region is dry, the thermal gradient follows the dry adiabat, which increases the temperature faster with depth than the moist adiabat. Because the moist region extends much deeper in the standard model, the temperature difference at depth will be even larger, as will be shown in later sections.

\subsection{Vapor cycle and energy budget}

\begin{figure}[htb]  
\includegraphics[scale=0.8,trim = 0.2cm 0cm 0.cm 0cm, clip]{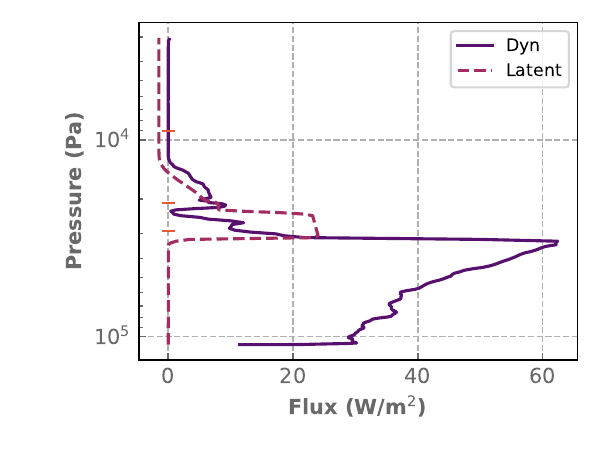}
\caption{Net upward energy fluxes due to sensible heat dynamical transport (solid) and latent heat (dashed). In the stable region, latent heat transported by vapor dominates the direct sensible heat flux due to turbulence, although the latter is not nil. See \app{app:nrj} for details of the calculation method.}
\label{fig:nrj_budget}
\end{figure}

We find that the thermal structure is strongly linked to the vapor cycle of the condensable species. The most apparent manifestations of this cycle in the simulations are the moist convective plumes (See \fig{fig:structure_moist}) that transport vapor upward in the moist troposphere. These plumes transport sensible and latent heat, as can be seen in \fig{fig:nrj_budget}. This upward motion is well-known for creating large amounts of condensates and precipitations. However, note that a sizable fraction of the vapor directly condenses at the top of the stable layer, forming a thin, horizontal cloud deck. This is evidenced by the sudden drop in the latent heat flux just at the top of the stable layer in \fig{fig:nrj_budget}.

Rainfalls then transport the condensable species back down in its condensed phase. The average rainfall rate can be estimated from the latent energy flux through the stable layer divided by the latent heat of vaporization and is about 10$^{-5}$kg/s.
As could be expected, the reevaporation of these rainfalls occurs below the base of the convective plumes. Indeed, a first requirement for reevaporation is to reach an unsaturated region below the updraft.

An interesting finding is that the level of reevaporation seems to set the bottom of the stable layer. Indeed, below this level, condensation is impossible, so there are no vapor sources/sinks. As a result, any vapor gradient in this lower region -- called dry troposphere in \fig{fig:structure_moist} -- would be short-lived and stellar energy deposited at depth is sufficient to trigger standard, dry convection and mix efficiently the atmosphere. 
But why cannot the dry troposphere extend above the reevaporation level? Let us remember that in a standard atmosphere, what drives convection in the first place is the fact that thermal radiation is not sufficient to carry the absorbed stellar radiation away, and convection stops at levels where this radiative cooling becomes efficient enough. Here, it takes a lot of energy to evaporate all the falling precipitations. This efficiently shuts down convection at the reevaporation level and energy is mostly carried upward by vapor in the form of latent heat above it (See \fig{fig:nrj_budget}). 

To close the cycle, vapor needs to find its way back toward the moist troposphere through the stable layer. This was a shortcoming of the model of \citet{LSH17}: they did not consider that the precipitations formed in the moist convective layer would necessarily reevaporate below the stable layer, so that the vapor would need to diffuse upward to maintain moist convection. In particular, this would have been in contradiction with their findings that no double-diffusive instability would develop in the stable layer, so that energy transport would be purely radiative. 

Our simulations shed a completely different light on this issue. Although we cannot test directly the "no double-diffusive instability" hypothesis because it would require a much finer resolution \citep{RGT11}, this issue is rendered rather moot by the other sources of turbulence in the system. 

\subsection{Turbulent mixing}

\begin{figure}[htb]  
\includegraphics[scale=0.8,trim = 0.2cm 0cm 0.cm 0cm, clip]{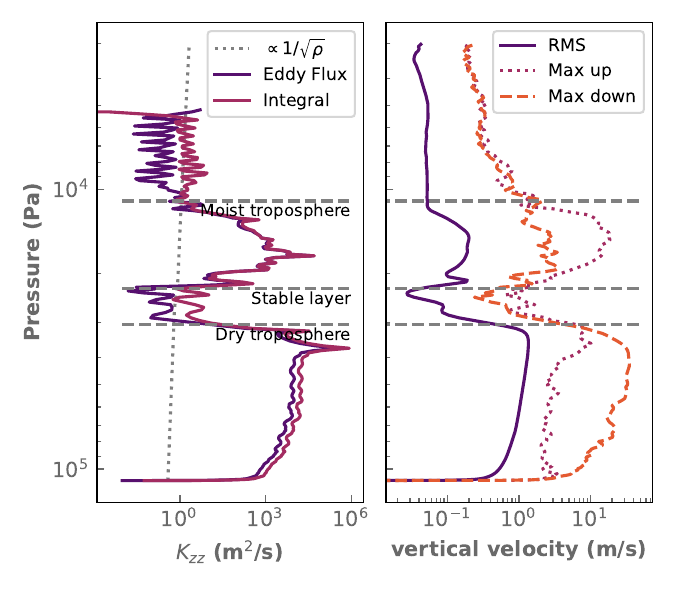}
\caption{\textbf{Left:} Vertical profile of the equivalent vertical mixing coefficient derived from the simulation using both the Eddy flux (\eq{eddyflux}) and Integral (\eq{integral}) methods . The dotted line shows the $1/\sqrt{\rho}$ trend for comparison. The horizontal lines depict the same levels as in \fig{fig:structure_moist}. \textbf{Right:} Profile of the vertical velocity showing the root-mean-square (solid), maximum upward (dotted) and downward velocity (dashed). Averages and maximum values are computed over temporal and horizontal dimensions.}
\label{fig:kzz}
\end{figure}

Even at the $\mathcal{O}(500\,m)$ scale resolved in our simulations, turbulence spontaneously appears\footnote{Turbulence here specifically refers to small-scale, but resolved motion that appears in stratified, stable regions.}. Even though the velocities involved are rather small (0.1\,m/s; see \fig{fig:kzz}), transport is still significant thanks to the steepness of the vapor and potential temperature gradient and the rather small vertical extent of the stable zone. 

To quantify this, we estimate the equivalent mixing coefficient (the so-called eddy diffusivity or $\kzz$) in the simulations using a passive tracer whose concentration ($\rtra$) is fixed at the surface and that undergoes advection by the flow and local exponential decay with a timescale $\tautra$. Then we use two different methods. In the first one, or \textit{eddy flux} approach, we just compute the average turbulent flux of tracer in the simulations and assume that:
\balign{
\kzz^{\mathrm{Eddy\,Flux}}\equiv \frac{\langle \rho \rtra \w \rangle}{ \rho\langle  \dpar{\rtra}{\z}\rangle }, \label{eddyflux}}
where $\langle \rangle$ denotes temporal and horizontal averaging. In the second method, or \textit{integral} approach, we use the fact that if the mixing exhibits a diffusive-like behavior, the steady-state tracer profile should obey the following law
\balign{\frac{1}{\rho}\dpar{\left(\rho \kzz \dpar{\rtra}{\z}\right)}{\z}=\frac{\rtra}{\tautra}.}
Integrating from the top of the atmosphere where we know that there is no vertical flux, one can get a second estimate
\balign{
\kzz^{\mathrm{Integral}}\equiv \frac{1}{\tautra }\frac{\int \rho \rtra \d \z}{\rho\langle  \dpar{\rtra}{\z}\rangle  }. \label{integral}}
Note in passing that because the transport is not perfectly diffusive (especially in convective regions), this approach can yield negative $\kzz$ so that we show the absolute value. With this important caveat in mind, both estimates (shown in \fig{fig:kzz}) exhibit a relatively good agreement and the differences inform us on the uncertainty that can be attributed to this parameter.

The profile of eddy diffusivities is fully consistent with the thermal profile found: stable layers exhibit low diffusivities whereas convective ones are more strongly mixed.
The values of the Eddy diffusivity in the stable regions is on the order of 0.1-10\,m.s$^{-2}$, which is comparable to values inferred for the lower stratosphere on Earth \citep{AYW81}. As we will discuss in more details in \sect{sec:1dmodel}, the vapor profile shown in \fig{fig:structure_moist} seems to favor eddy diffusivities that are at the lower end of this range and decreasing slightly with altitude in the stable region. This is evidenced by the vapor gradient steepening with altitude as we go from the bottom to the top of the stable layer.  Indeed, 
there is no condensation or reevaporation in that region so that, in steady state, the upward vapor flux ($\propto\kzz(\d \q / \d z)$) needs to be constant. So a steepening gradient means a decreasing mixing.

This turbulence seems to be produced by the dry updrafts hitting the bottom of the stable layer \citep{LSC03}. This both creates upward propagating gravity waves and directly stirs the medium. We believe that gravity waves themselves do not participate much in the mixing in the stable region  because i) they do not transport matter in the linear regime and ii) their amplitude is too low at this level to break and induce subsequent mixing. 

\subsection{Dynamics}

The velocity distributions shown in the right panel of \fig{fig:kzz} are very different between the two convective zones. In the bottom, dry convective region, we observe a permanent overturning circulation with velocities on the order of 1\,m/s. At this depth, this is largely driven by stellar radiation that still penetrates efficiently.
On the contrary, in the moist convective region, there is very little background motion apart from upward propagating gravity waves that still have a low amplitude as they are not far from their launching region (a few 10$^{-1}$\,m/s). The only exception occurs in the thin boundary layer just above the stable layer below, around $2\times10^4$\,Pa. We can see a local maximum of both the RMS velocity and the eddy diffusivity in \fig{fig:kzz}. In this region, the atmosphere is locally unstably stratified, driving small-scale dry convection which transports humidity upward until saturation induces moist convection.

Then, as can be seen in \fig{fig:structure_moist}, moist convective plumes form episodically to release the energy and vapor that has built-up at the top of the boundary layer. As is common on Earth, these moist convective plumes are much narrower and reach faster speeds than their dry counterparts. As also seen on Earth, the rising plumes are mirrored by cold downdraft created by reevaporation below the convective cloud: the cold pools. One big difference is that on Earth, the moist updrafts reach higher velocities than the cool downdrafts. This is because in our atmosphere, updrafts are powered by the combined power of latent heat and compositional buoyancy (remember that water vapor is lighter than molecular nitrogen) whereas these two effects compete in downdrafts, with the thermal effect of latent heat dominating.

In a hydrogen-dominated atmosphere like K2-18\,b's, vapor is usually heavier than dry air. So latent heat release by condensation needs to fight against the stabilizing effect of the mean molecular weight gradient to power the rising, moist plumes. In fact, it can win only when the vapor concentration is below the critical ratio, which is the very reason why there is a stable region in the first place. But even in the moist region, this competition leads to relatively sluggish upward motion. In comparison, the downdrafts that form below the rising plumes are much more vigorous, with a factor $\times \,$2-3 in velocity, because the vapor loading helps the cooling due to reevaporation (that can reach several K) in accelerating the plume downward. 

Finally, one can see chevron-like structures around the rising plume in the third panel of \fig{fig:structure_moist}. These are typical of convection induced gravity waves that can propagate as the moist region is stably stratified (in the sense that it has a positive Brunt-Väisälä frequency).

\subsection{Stabilisation and subsaturation}

An intriguing feature of our simulations is that the atmosphere is subsaturated almost everywhere but for a thin region near the top of the stable layer and in the core of ascending moist plumes. Yet, the atmosphere exhibits a stable layer where vapor concentration exceeds the critical inhibition fraction as predicted by \citet{LSH17}. These two statements seem in contradiction as the analytical theory invokes saturation to suppress convection. However, as discussed in Sect 3.4 of \citet{LSH17}, saturation is invoked only to suppress \textit{moist} convection, and saturation indeed needs to occur to form moist convective updrafts as visible in the third panel of \fig{fig:structure_moist}.

However, the medium does not need to be saturated for the vapor gradient to have a stabilizing effect---thermohaline convection being a perfect example \citep{Led47,Ste60}. So saturation needs to happen \textit{somewhere} in the convective parts of the atmosphere to drive the vertical vapor concentration gradient between a low value above the cloud deck and a high internal value. But our simulations confirm that the stable layer can be largely subsaturated and remain stable. 

Interestingly, we find that the saturated layer that coincides with the top of the stable layer happens exactly at the critical concentration value. Again, this makes sense because above that level, moist convection can occur. Moist convective regions must be subsaturated on average (see \fig{fig:structure_moist}) because of the dry subsidence regions that appear to compensate for the upward mass flux in the convective plumes. In the stable region, where there is very little motion, the vapor concentration is much more horizontally and temporally uniform. Hence it is always close to saturation at the layer top because this is where moist plumes originate.

\section{A simplified 1D framework for fast modeling}\label{sec:1dmodel}

Our 3D simulations are too expensive to be carried over a large diversity of conditions. Now that we have outlined the most important features of the structure of Neptune-like atmospheres, we present a 1D model that is able to reproduce these features for only a tiny fraction of the computational burden. This model is based on the \texttt{Exo\_k} library \citep{Lec21}\footnote{\url{https://perso.astrophy.u-bordeaux.fr/~jleconte/exo_k-doc/index.html}} that has been recently updated with a full-fledged time-stepping atmospheric evolution package described in \citet{SLT23}.

This atmospheric evolution package has the advantage of being extremely flexible while using some computational tricks to remain very fast. Hereafter, we will focus mainly on the new features of the model that pertain to the inhibition of convection.

\begin{figure*}[htb]  
\includegraphics[scale=1.,trim = 0.2cm 0cm 0.cm 0cm, clip]{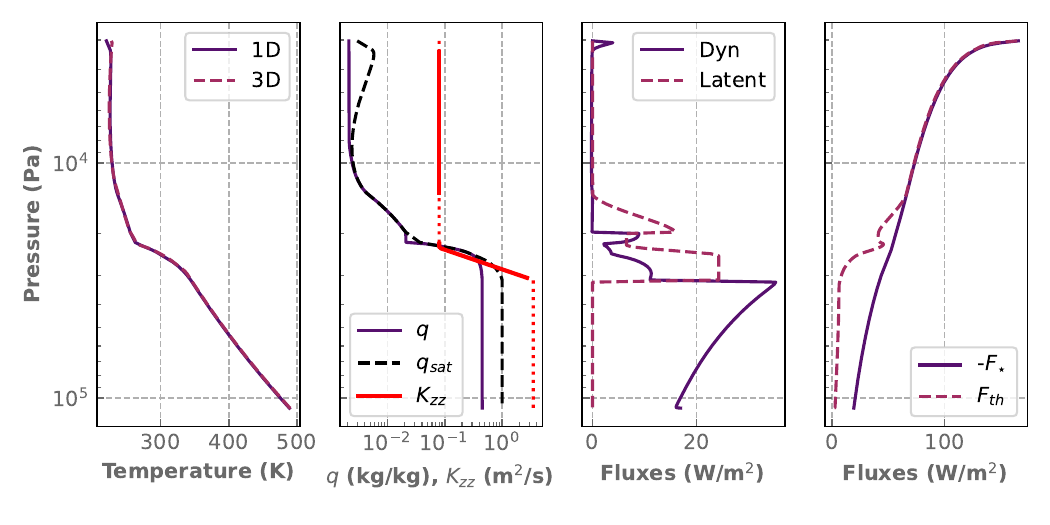}
\caption{Atmospheric structure of the 1D model for the baseline case. From left to right: Temperature, Water vapor specific concentration, dynamical fluxes, and radiative fluxes vertical profiles. The dashed curve in the temperature panel shows the average 3D thermal profile for comparison. In the second panel, the dashed line shows the value of the saturation concentration ratio and the red curve shows the profile of eddy diffusivity. Dots show convective zone where mixing is not diffusive so that the value of the diffusivity is irrelevant there.  The third panel can be compared to \fig{fig:nrj_budget}. The last panel shows (minus) the net upward incoming stellar flux (solid) and the net outgoing thermal flux emitted by the atmosphere (dashed). This latter panel shows that most of the radiative cooling to space occurs in the stable layer and above.}
\label{fig:1D_structure}
\end{figure*} 

\subsection{Criteria for convection inhibition}

To allow for relatively large timesteps, \texttt{Exo\_k} treats both moist and dry convection using standard adjustment schemes that identify sets of adjacent layers that are unstable and compute the energy fluxes necessary to bring back these layers to the relevant adiabat. 

For dry convection, unstable layers were identified by regions of decreasing potential temperature ($\theta$). This is equivalent to the \citet{SH58} criterion. When one wants to account for the mean molecular weight effect of the vapor, one needs to use the \citet{Led47} criterion instead. But as is well known in terrestrial meteorology (see also \citealt{LSH17}), this can simply be recast as identifying regions of decreasing \emph{virtual} potential temperature,
\balign{\tetav\equiv\left(1-\frac{\Mv-\Md}{\Mv} \qv\right) \  \teta,}
inducing minimal changes to the code. Any set of unstable layers is brought back to a neutral state with a uniform composition and potential temperature (hence uniform virtual potential temperature) in a single timestep and conserving the total enthalpy of the layer. Layers above or below the newly adjusted layer that have been destabilized by the adjustment are themselves adjusted iteratively. This treatment is a bit simpler than the one proposed in \citet{HP23} for dry compositional convection. Our scheme nonetheless performs very well in the bottom dry-convective region, as shown hereafter. This is probably due to the fact that mixing in this layer is continuously driven by thermal and compositional effects and is thus relatively efficient.

In the previous version of \texttt{Exo\_k} \citep{SLT23}, moist convective adjustment was triggered when the temperature gradient was larger than the moist adiabat and when the medium was saturated \citep{MW67}. Following the analysis in \citet{LSH17}, moist-convection inhibition has been accounted for by simply suppressing the adjustment in any layer where $\qv > \qcri$. In the convective regions, the excess vapor condensed during the adjustment process is assumed to instantaneously rain down to the reevaporation level.

In essence, these changes are sufficient to naturally force a stable layer in our unidimensional model. This however requires the ability to trace the cycle of both the vapor and the condensates. For this we use two tracers that are mixed by convection as described above. For the thermodynamics and the microphysics of clouds and precipitations, we use exactly the same parametrization as in our 3D model, which is described in \sect{sec:phys_param}, ensuring that any difference will be due to the dynamics.

\subsection{Turbulent mixing in stable layers}

While it is not needed to create a stable layer, we have seen in \sect{sec:3dmodel} that turbulent mixing is an important ingredient in determining the strength of the vapor cycle and, to a lesser extent, in transporting sensible heat through the stable layer. To incorporate that, we added a diffusive flux of tracers and entropy of the form $\kzz(\d  / \d z)$, where the eddy diffusivity, $\kzz$, is a free parameter. In our simplest model, hereafter called the $\textit{constant diffusivity}$ case, this parameter is constant throughout the atmosphere and calibrated to yield the proper flux of vapor through the stable layer. This yields a value of about $0.3$\,m$^2$/s, which is representative of the values found in the stable layer in \fig{fig:kzz}. Remember that tracers are already fully mixed in the convective regions, which emulates an infinite eddy diffusivity. So there is no need to increase our $\kzz$ there.

Yet, we find that if one wants to model very accurately the shape of the thermal profile in the stable zone, a constant $\kzz$ provokes too abrupt a transition at the top of the dry convective region, where convective plumes can overshoot. To model this, we implement an alternative formulation, referred to as the \textit{baseline} scenario, where the eddy diffusivity assumes a larger value ($\kzzmax$) just at the top boundary of the dry convective region (also called the radiative-convective boundary, whose pressure is $\prcb$) and drops off very rapidly as some high power $\alpha$ of the pressure before settling to a lower constant value ($\kzzmin$) higher up: 

\balign{
\kzz = 
\left\{\begin{array}{ll}
\kzzmax & \press > \prcb
\\
 \mathrm{Max}(\kzzmax (\press/\prcb)^\alpha, \kzzmin) & \press < \prcb.
 \end{array}\right.
}
The value of $\alpha$ is set to reproduce the sharp decrease of the turbulence above the convective region seen in \fig{fig:kzz}. This yields $\alpha\approx13$. Then, $\kzzmin$ and  $\kzzmax$ are tuned to reproduce the vapor flux and the thermal gradient just above the radiative-convective boundary, respectively. This yields $\kzzmin\approx 0.08$\,m$^2$/s and  $\kzzmax\approx 3$\,m$^2$/s, which is also consistent with the numerical values from the 3D simulation.

\subsection{1D/3D comparison}

In \fig{fig:1D_structure}, we present the atmospheric structure for the \textit{baseline} scenario. Notice that because there are some energy losses in the dynamics of the 3D model (see \sect{sec:nrjcons}), the baseline case is ran with a decrease of the incoming stellar flux of the corresponding amount to allow for a proper comparison. 

The agreement between the 1D and 3D models is rather striking. With very few free parameters, not only both thermal structures are very close, but the shape and magnitude of the various energy fluxes are reproduced as well. We even recover the thin, dry boundary layer between the top of the stable layer and the moist convective layer, which is evidenced by the small layer with constant vapor concentration near $2\times10^4\,$Pa in the second panel of \fig{fig:1D_structure}. This shows that this boundary layer is not created by dynamical requirements alone. The most notable discrepancy is the fact that the moist convective layer is fully saturated in our 1D model, which is to be expected because saturation is a prerequisite to the onset of convection in our moist adjustment scheme. Our 3D simulations are closer to what happens in reality where saturation is only required in the rising plumes and dry subsident regions force the convective region to be subsaturated on average.

\begin{figure}[htb]  
\includegraphics[scale=1.,trim = 0.2cm 0cm 0.cm 0cm, clip]{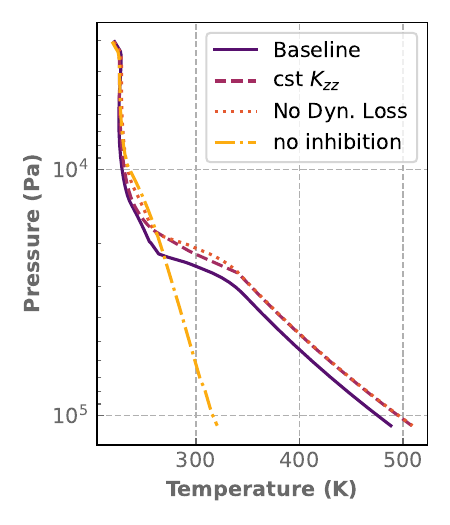}
\caption{Sensitivity analysis showing the impact of various assumptions on the equilibrated temperature profile. The solid curve is the baseline profiles, the dashed one shows the case with a constant eddy diffusivity, and the dotted one shows the effect of correcting for the dynamical losses. Differences between the various cases is much smaller than the effect of convection inhibition itself which can be seen by comparing with the no-inhibition case (dash-dotted curve). }
\label{fig:sensitivity}
\end{figure}

\fig{fig:sensitivity} shows how the thermal profile is modified when changing the profile of the turbulent eddy diffusivity. As expected, the change in the slope of the thermal gradient is more abrupt at the top of the dry convective zone in the case of a constant eddy diffusivity. This changes the 1 bar temperature by about 20\,K. We also ran the baseline model but with the same actual input flux as the 3D simulations to quantify the potential effect of the dynamical losses of the latter. This also causes an increase of the deep adiabat of about 20\,K. Although not negligible, we have to bear in mind that these differences are much smaller than the effect of the inhibition itself, which raises the temperature at the 1\,bar level by $\sim$200\,K.

%
%

\section{Observational markers of convection inhibition}\label{sec:observability}

As can be seen in \fig{fig:sensitivity}, the presence or absence of a stable layer (i.e. of convection inhibition) affects relatively mildly the temperature of the upper troposphere. This is because the atmosphere above the stable layer is already optically thick in most of the thermal part of the spectrum. As a result, the temperature of the atmosphere below the stable layer affects only slightly the outgoing flux and only a small change in the temperature of the photosphere is needed to reach global radiative equilibrium.

To identify a more robust marker of convection inhibition in temperate sub-Neptunes, we turn to the composition of the atmosphere. Indeed, we know that the chemical composition of temperate atmospheres connected to a deep gaseous envelope is mainly determined by the temperature of the level at which chemical processes are quenched \citep{VDM18}. Above the quench level, chemical reactions are too slow compared to the mixing by the atmosphere. Above that level, the composition is thus rather uniform up to the level where photochemical rates start to dominate. For a given elemental abundance, the molecular content of the atmosphere can therefore be a tracer of the deep temperature. 

\begin{figure*}[htb]  
\sidecaption
\includegraphics[scale=1.,trim = 0.cm 0cm 0.cm 0cm, clip]{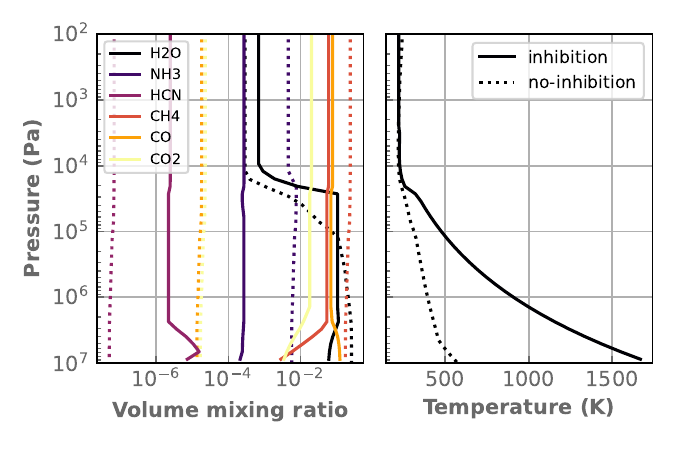}
\caption{Vertical profiles of molecular volume mixing ratios in the atmosphere of our prototypical temperate Neptune with (solid) and without (dotted) convection inhibition. The metallicity is assumed to be 300$\times$solar. The right panel shows the thermal profiles used for the two cases, where we see that the moist troposphere extends quite deep in the no-inhibition case, resulting in a much lower quenching temperature. }
\label{fig:chemistry}
\end{figure*}

The main implication of the presence of a stable layer is the higher temperature in the deep atmosphere. This is particularly visible in the right panel of \fig{fig:chemistry} where we have extended our 1D model to 1000\,bars. The increase in temperature is due to two effects: i) the temperature jump in the stable layer and ii) the fact that the dry troposphere, which has a much steeper lapse rate, starts much higher. Hence, differences can be up to 1000\,K at the 100\,bar level.

To quantify the effect on the chemistry, we have computed the chemical composition of the two model atmospheres in \fig{fig:chemistry} using the chemistry module of Exo-REM \citep{BCB21} assuming $\kzz=10^4\,$m$^2$/s. We take this value as characteristic of the deep convective region where the quenching occurs. Our tests show a rather low sensitivity to the exact value of this parameter. Exo-REM computes the quenching levels and resulting disequilibrium composition by comparing the mixing timescale with the chemical timescale using formulas from \citet{ZM14}. This yields a much lower quenching pressure of $\approx 20\,$bar for the baseline scenario compared to $\approx 600\,$bar for the no-inhibition case.

As can be seen in the left panel of \fig{fig:chemistry}, the two cases predict very different chemical abundances in the stratosphere. Notably, the no-inhibition case - with its low temperature interior - exhibits very low levels of CO and CO$_2$ in the stratosphere, most of the carbon forming CH$_4$. On the contrary the case where inhibition is taken into account, CO, CO$_2$, and CH$_4$ are all in detectable quantities. This case also predicts a much lower abundance of NH$_3$. These conclusions are on par with the conclusions of \citet{CVS17} for Uranus and Neptune.

As water vapor is cold-trapped at the tropopause in all cases, carbon bearing species dominate the transit spectrum. The two model atmospheres described above would thus be easily distinguishable observationally with JWST. Observing a Neptune-like planet receiving an insolation below the runaway greenhouse threshold would thus enable us to infer the presence of convection inhibition and a stable layer.

\section{Atmospheric constraints on the existence of liquid oceans on K2-18\,b} \label{sec:runaway}

It has been recently claimed that the absence of ammonia in the transit spectrum of K2-18\,b could be the sign of the existence of a liquid water ocean below a relatively shallow atmosphere of less than a few bars \citep{MSC23}.

The main argument against such a scenario is that the irradiation level received by the planet is above the critical \textit{runaway greenhouse} threshold for cloudless H$_2$ rich atmospheres \citep{ITP23}. But \citet{MSC23} argues that tropospheric clouds or hazes could increase the albedo of the planet, stabilizing the liquid ocean.

In this section, we revisit these arguments using our newly developped 1D atmospheric model and the improved knowledge of the atmospheric composition of the planet provided by the recent JWST observations of the system \citep{MSC23}. In particular we show that
\begin{itemize}
\item[$\bullet$] When convection inhibition is accounted for, the planetary albedo required to stabilize an ocean on K2-18\,b is higher than previously estimated,
\item[$\bullet$] Unlike on solar system planets, tropospheric clouds cannot provide such high albedos because a significant part of the stellar flux does not reach the troposphere to be reflected,
\item[$\bullet$] When we add sufficient levels of stratospheric haze to our model atmospheres to reach the albedos necessary to keep a liquid ocean, the methane features in the transmission spectrum are too muted to be consistent with recent observations. 
\end{itemize}

\subsection{Runaway greenhouse threshold and convection inhibition}

Our first goal is to assess the maximum stellar irradiation K2-18\,b can receive while still sustaining a liquid ocean below the H$_2$ dominated atmosphere. This question has been investigated by \citet{ITP23}, who showed that the greenhouse effect of an H$_2$ dominated atmosphere is much greater than that of an N$_2$ dominated atmosphere, strongly lowering the runaway greenhouse threshold. We revisit here these calculations for several reasons: i) we want to use the exact planetary parameters of K2-18\,b, ii) JWST data now provide an estimate of the atmospheric composition of the planet, which allows for more accurate opacity and mean molecular weight estimates, and iii) our 3D simulations provided a better understanding of the turbulent transport mechanisms in those atmospheres, whereas \citet{ITP23} assumed that energy was solely transported by radiation in the stable layers of the atmosphere.

Because the irradiation received by the planet is rather well constrained, we reframe the issue by asking what would the planetary Bond albedo need to be for the absorbed insolation to remain below the runaway greenhouse threshold. To answer this question, we perform the following experiment : we equilibrate our 1D model for a given surface pressure with a stellar irradiation equal to $\Finc (1-\alb)$ where $\Finc = 342\,$W/m$^2$ is the average insolation received by K2-18\,b and $\alb$ is an albedo that we choose arbitrarily. In this specific experiment, we do not include any aerosols in the atmospheric model as their effect is wholly incorporated in the $\alb$ parameter. Because K2-18 is a red star and that significant amounts of methane have been detected in the planet's atmosphere, almost all the stellar light arriving at the atmospheric top is absorbed in this setup so that $\alb$ is a good proxy for the bond albedo of the model.

There are two differences with respect to 1D simulations shown in the previous sections. First, we recomputed opacity tables with mixing ratios of methane and carbon dioxide of 10$^{-2}$ppmv, which seem to be the best match for the JWST observations of \citet{MSC23}. Second, we change the surface boundary condition to better mimic an ocean. Instead of fixing the mixing ratio of vapor in the lowest layer equal to an expected value at depth, we now treat the surface as an infinite source of water and vapor can freely evaporate in the first layer until saturation is met. As in \citet{ITP23}, we keep the \textit{total} surface pressure constant during the evolution. We then let the model evolve until it either reaches thermal equilibrium, in which case we deem the ocean stable, or the mixing ratio of vapor reaches unity at the surface, which we take as a proxy for the onset of runaway greenhouse. 

\begin{figure}[htb]  
\includegraphics[scale=0.85,trim = 0.cm 0cm 0.cm 0cm, clip]{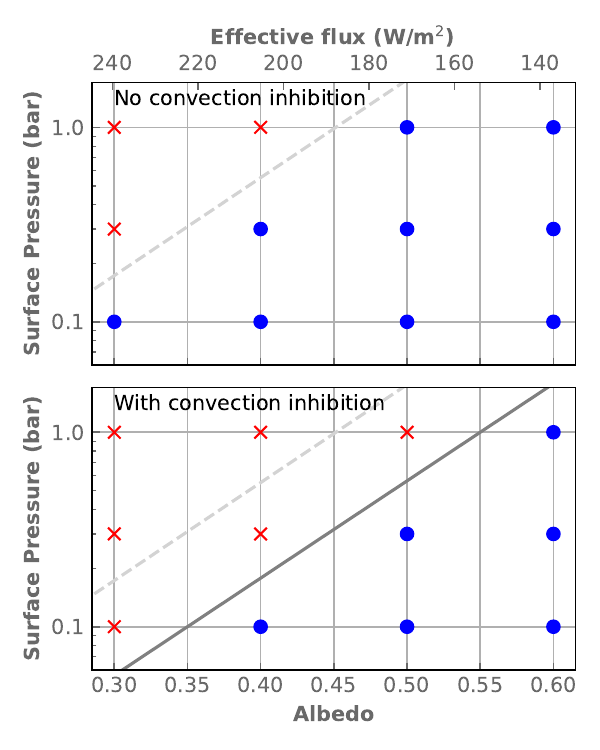}
\caption{Constraints for the presence of a liquid surface ocean. Each marker shows the outcome of a simulation for a given imposed albedo (the equivalent effective flux is shown on the top axis) and atmospheric surface pressure. Blue dots show cases where a liquid surface ocean is stable. Red crosses show cases where a steam atmosphere forms. The top panel shows the results of traditional models where convection inhibition is disregarded and the bottom panel shows results with convection inhibition. The dashed (solid) line roughly depicts the limit to the stability of an ocean in the case without (with) inhibition. One can see that the inhibition limits the stability of oceans to higher albedos, i.e. less irradiated planets.}
\label{fig:ocean_limits}
\end{figure}

The results of this experiment are summarized in \fig{fig:ocean_limits}. As expected, the higher the surface pressure, the higher the Bond albedo needs to be to keep a stable surface ocean. These results are in rough agreement with the results of \citet{ITP23}. First, we confirm that convection inhibition decreases the threshold for the onset of runaway greenhouse. This is because near the critical insolation, the moisture always becomes sufficient near the surface to shut down moist convection, increasing the surface temperature for a given outgoing flux, thus increasing the greenhouse effect of the atmosphere. Second, \citet{ITP23} find that the maximum stable insolation for a 1\,bar atmosphere around an M star is around 110\,W/m$^2$ whereas our last stable insolation is $\approx$\,130\,W/m$^2$ ($\alb=0.6$). The discrepancy could be partly due to the presence of methane, whose anti-greenhouse effect can be rather strong around late-type stars, but we think that the main difference is our treatment of the turbulent transport in the stable layer where convection inhibition operates. Because turbulence transports both sensible and latent heat, the thermal gradient is much less steep in our model compared to a fully radiative zone, which weakens the greenhouse effect of the stable layer. We verify this by re-running this case with a much weaker turbulent transport and find that it indeed enters a runaway greenhouse phase. Further comparison is however difficult as the model currently relies on turbulence to transport water vapor in non-convective zones so that removing it entirely causes the stable layer to become unsaturated. But this shows that accounting correctly for the dynamics of the atmosphere is important to derive quantitative limits. 

However, the limits we find are much more stringent than the ones found by \citet{MPC21}. For an M star like K2-18, they find that the maximum equilibrium temperature (that is corrected for the Bond albedo) to keep a liquid surface ocean is $\approx$\,410\,K --- which corresponds to a planet averaged absorbed/thermally emitted flux of $\approx$\,1300\,W/m$^2$. This is to be compared to our limit for the 1\,bar case, which is estimated to be $\approx$\,230\,K (150\,W/m$^2$). To put this into context, the current equilibrium temperature of the Earth is $\approx$\,255\,K (240\,W/m$^2$), and recent estimates of the runaway greenhouse limit for Earth-like planets yield estimates between $\approx$\,260 and 270\,K (270-300\,W/m$^2$), depending on the treatment of continuum opacities, clouds, and atmospheric dynamics \citep{KRK13, LFC13, YLW16}. The fact that the limit for hydrogen-dominated atmospheres occurs at lower fluxes is due to the increased greenhouse effect of H$_2$ compared to N$_2$, and has been extensively studied \citep{KC19, CTB22, ITP23}. 

The reason that \citet{MPC21} find such high limits is less clear. It seems to be due to their use of an \textit{ad hoc} -- and rather extreme -- approximation to treat aerosols: they assume that aerosols can be arbitrarily efficient scatterers and model them by multiplying the Rayleigh scattering coefficient of H$_2$ by an arbitrarily large factor until the atmospheric Bond albedo reaches the desired value. In addition to increasing the albedo, this causes the stellar radiation to be scattered many times in the stratosphere, which, counter-intuitively, enhances absorption there. Around redder stars, this results in stratospheres that are as hot, if not hotter, than the surface, which effectively suppresses the greenhouse effect of all the atmospheric gases.

However, as we will see in the next section, the presence of such reflective haze particles in the stratosphere is contradicted by observational data.

\subsection{Constraints on aerosols}

In this section, we make an attempt at better quantifying the limits that can be put on the albedo that aerosols (either clouds or hazes) can realistically produce on a planet like K2-18\,b. 

A first hypothesis put forward by \citet{MSC23} is that the presence of deep, highly reflective tropospheric clouds could produce a sufficient albedo to stabilize an ocean. Although such clouds are able to produce high albedos for our solar system giant planets, we have to remember that K2-18 is an M dwarf with an effective temperature around 3500\,K and that its radiation is easily absorbed in the stratosphere of the planet by the multiple near-infrared methane bands. In a cloudless model of K2-18\,b produced with the fiducial methane and carbon dioxide mixing ratios of 10$^{-2}$\,ppmv found by \citet{MSC23}, half the flux is absorbed above the 100\,mb level, which is still higher than the tropopause. So no scattering happening below this level, however intense, could increase the albedo above 0.5 (and that does not even account for the fact that scattered light would have to cross the stratosphere a second time to escape). This is well illustrated by the 3D global climate models from \citet{CBB21} who found that the albedos of their models for K2-18\,b barely exceed 0.12 even when thick dayside tropospheric water clouds form.

Another hypothesis is the presence of highly reflective haze in the stratosphere, although neither \citet{MPC21} or \citet{MSC23} discuss which type of haze could meet the required constraints. This solution works in principle because it is able to scatter incoming stellar light high in the stratosphere, before it is efficiently absorbed. However, it is easy to see that such a reflective haze should also strongly affect (e.g. flatten) the transit spectrum of the planet in the visible and near-infrared, whereas the recent JWST spectrum of \citet{MSC23} shows relatively deep methane absorption features with an amplitude in excess of 100\,ppm between 1 and 1.5 micron. To quantify this effect, we compute eclipse and transmission spectra of the fiducial model of K2-18\,b discussed above where we add haze scattering following the parametrized approach of \citet{MPC21}. In this approach, the amount of haze is encompassed in a so-called haze factor ($n_{haze}$), which is used to multiply the cross section of Rayleigh scattering of H$_2$. Let us note that $n_{haze}=1$ corresponds to the fiducial clear atmosphere model.

\begin{figure}[htb]  
\includegraphics[scale=0.67,trim = 0.cm 0cm 0.cm 0cm, clip]{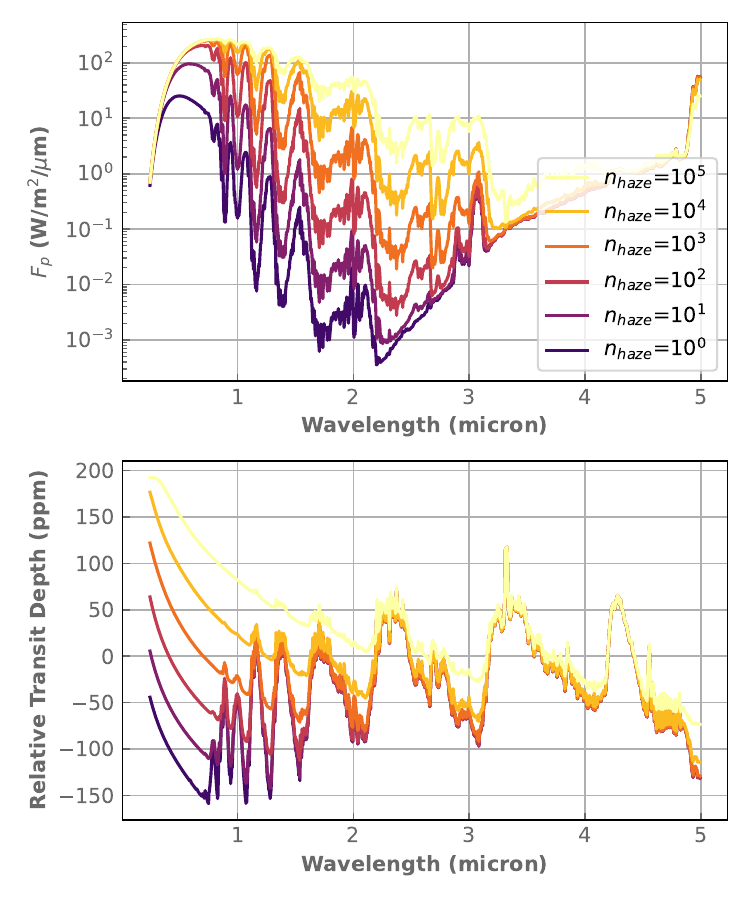}
\caption{Eclipse (top) and transit (bottom) spectra of our models of K2-18\,b with hazes parametrized through the $n_{haze}$ factor (see text). One can see that when the amount of haze increases, the amount of reflected light (hence the albedo) increases but the amplitude of the methane bands in the transit spectrum decreases. }
\label{fig:haze_factor_spectra}
\end{figure} 

\begin{figure}[htb]  
\includegraphics[scale=0.8,trim = 0.cm 0cm 0.cm 0cm, clip]{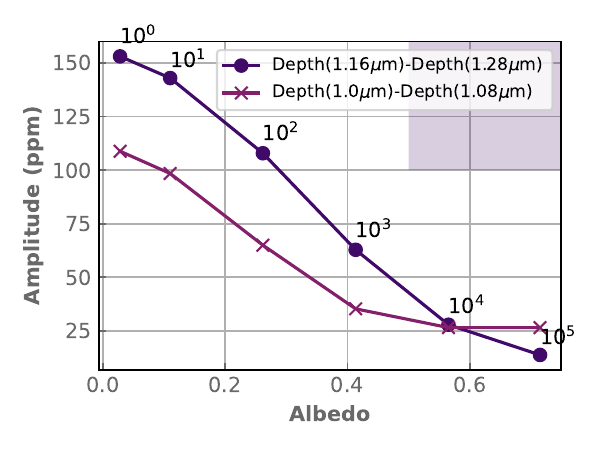}
\caption{Amplitude of two near-infrared methane molecular bands in the transit spectrum of our models of K2-18\,b with parametrized hazes as a function of the Bond albedo. As already illustrated in \fig{fig:haze_factor_spectra}, the amplitude of the methane bands decreases when the albedo of the planet increases. The numbers show the values of $n_{haze}$. The shaded area roughly depicts the area of parameter space that could be consistent with both the observations (transit amplitude greater than 100\,ppm) and the albedo required to sustain a surface ocean ($\alb \gtrsim 0.5$). The relatively high band amplitude found in the observations rules out haze that are reflective enough to stabilize a liquid surface ocean.}
\label{fig:haze_factor}
\end{figure}

The resulting spectra are shown in \fig{fig:haze_factor_spectra}, where we see that the amount of reflected light increases with $n_{haze}$, as expected. The corresponding albedo can be seen in \fig{fig:haze_factor}, with $\alb=0.03$ for the clear case and $\alb=0.72$ for $n_{haze}=10^5$. However, one can see in the bottom panel of \fig{fig:haze_factor_spectra} that the increased scattering also mechanically suppresses the molecular methane features in the near-infrared. We quantify the amplitude of these features by taking the difference between the transit depth at two absorption peaks (1. and 1.16$\mu$m) and the depth in the nearest windows (1.08 and 1.28$\mu$m respectively). Those amplitudes are shown as a function of the model albedo in \fig{fig:haze_factor}. As could be expected, the amplitude of the molecular features decreases when the albedo increases. But more importantly, we cannot find a model where we have both a sufficient albedo and a transit amplitude that matches the data.

Because the aforementioned parametrization of haze is rather \textit{ad-hoc}, we have tested various types of aerosols, including water ice particles and venusian sulfuric acid cloud particles that are known for their high reflectivity. Although we do not show all the results here, we always find a very similar trend to what is shown in \figs{fig:haze_factor_spectra}{fig:haze_factor}: models with high albedos produce very flat spectra that do not match the observations. And it is difficult to imagine a species that would reflect enough light in an unobserved part of the spectrum because the NIRISS SOSS observation precisely cover the peak of the stellar emission.

So we conclude that the observations of K2-18\,b make the possibility of a planet harboring a liquid ocean thanks to a haze-driven high-albedo very unlikely. The only remaining possibility would be that hazes would form only on the dayside to almost disappear at the limbs. However, the reader should bear in mind that, unlike clouds, hazes cannot easily sublimate and thus usually are much more uniformly distributed than clouds -- as shown by solar system examples like Titan or Giant Planets. We thus deem this possibility rather unlikely as well.

\section{Conclusions}

We have developed a cloud-resolving model able to simulate moist-convection in vapor enriched atmosphere. Being integrated in the ecosystem of the Generic PCM model, it is very flexible and can be easily adapted to a wide diversity of planets. We then have investigated how moist convection behaves in H$_2$ dominated atmospheres using K2-18\,b as a prototypical temperate Neptune-like planet.

Our main general findings are that
\begin{itemize}
\item[$\bullet$] Moist convection is effectively inhibited when the vapor abundance exceeds the threshold abundance given by linear theory \citep{Gui95, LSH17}
\item[$\bullet$] The atmospheric structure envisioned by \citet{LSH17} -- i.e. the formation of a stable layer between a moist troposphere above and a dry troposphere below -- is recovered in the 3D simulations, even though some of the fundamental assumptions of the analytical theory are not verified. In particular, almost no part of the atmosphere is fully saturated. 
\item[$\bullet$] The stable layer harbors some turbulence, the magnitude of which controls the intensity of the vapor cycle in the atmosphere. Both the latent and sensible heat transport that result contribute significantly to the energy flux through the stable layer and need to be accounted for when determining the thermal structure.
\item[$\bullet$] The deep gaseous envelope of Neptune-like planets where condensation occurs should be much hotter than envisioned by standard models. This impacts the chemical composition of the atmosphere. This could itself be a way to ascertain the presence of a stable layer in the atmosphere. 
\item[$\bullet$] The higher temperature at depth also decreases the limiting insolation at which a surface liquid ocean can be stable under a H$_2$ dominated atmosphere (also see \citealt{ITP23}). 
\end{itemize}

Although our conclusions are based on simulations that focus on the water condensation region, the principles are rather general and should readily apply to the condensation level of any other condensable species that is abundant enough: methane in Uranus and Neptune \citep{CLS23}, iron or silicates near the core of Neptune-like planets \citep{MGS22,MS22}, etc.


The fact that the higher temperatures at depth further reduce the insolation at the so-called \textit{inner edge of the habitable zone} for H$_2$ dominated atmospheres has direct implications for observed planets that are in this range of insolations. In particular, the insolation received by K2-18\,b seems too high to find a configuration that could explain both the very high albedo necessary to stabilize a surface ocean under its H$_2$ atmosphere and the rather deep methane features observed in transmission spectroscopy. Therefore, if the non-detection of ammonia in this atmosphere is confirmed, it might be necessary to invoke other mechanisms for the lack of this molecule than a shallow atmosphere above a liquid ocean, like

\begin{acknowledgements}
The authors acknowledge the support of the French Agence Nationale de la Recherche (ANR), under grant ANR-20- CE49-0009 (project SOUND), the Programme National de Plan\'etologie (PNP), and CNES. ML acknowledges funding from the European Union's Horizon Europe research and innovation program under the Marie Sk\l odowska-Curie grant agreement "MuSICA-V"
\end{acknowledgements}

\bibliographystyle{aa}
\bibliography{biblio}

\newpage

\begin{appendix}

\section{Calculation of the energy budget}\label{app:nrj}

To help visualize the flow of energy in our simulations, we compute the net upward energy fluxes as follows. In the 3D model, the physical parametrizations give us at each timestep the specific heating rate due any diabatic heating process in each cell ($\Qi$).
The net upward flux for process $\i$ at any level defined by the hydrostatic pressure $\hpm$ is defined as the integral of the heating between that level and the surface
\balign{
\Fluxi = -\int_\hpm^{\hpms}\Qi\frac{\d \hpm}{\grav},
}
where the minus sign comes from the ordering of the integral boundaries. 

The sensible heat transport is computed from the velocity and potential temperature fields using $\Flux_\mathrm{dyn}\equiv \cp \langle\rho \w' \teta ' \rangle$, where the primes denote departures from the average values.

\end{appendix}


\end{document}